\documentclass[preprint,
               superscriptaddress,preprintnumbers,amsmath,amssymb,pra]{revtex4}
\usepackage{graphicx}
\usepackage{amsmath}

\begin{document}
\newcommand{\ve}{\varepsilon}
\newcommand{\vel}{\varepsilon_l}
\newcommand{\peq}{P_{\rm eq}}
\newcommand{\tpeq}{\widetilde{P}_{\rm eq}}
\newcommand{\pneq}{P_{\rm neq}}
\newcommand{\kb}{{k_{\bot}}}

\thispagestyle{empty}

\title{Nonequilibrium effects in the Casimir force between two similar
metallic plates kept at different temperatures}

\author{G.-L.~Ingold}
\affiliation{Universit\"{a}t Augsburg, Institut f\"{u}r Physik,
86135 Augsburg, Germany}

\author{
G.~L.~Klimchitskaya}
\affiliation{Central Astronomical Observatory at Pulkovo of the
Russian Academy of Sciences, Saint Petersburg,
196140, Russia}
\affiliation{Institute of Physics, Nanotechnology and
Telecommunications, Peter the Great Saint Petersburg
Polytechnic University, Saint Petersburg, 195251, Russia}

\author{
V.~M.~Mostepanenko}
\affiliation{Central Astronomical Observatory at Pulkovo of the
Russian Academy of Sciences, Saint Petersburg,
196140, Russia}
\affiliation{Institute of Physics, Nanotechnology and
Telecommunications, Peter the Great Saint Petersburg
Polytechnic University, Saint Petersburg, 195251, Russia}
\affiliation{Kazan Federal University, Kazan, 420008, Russia}

\begin{abstract}
We study the Casimir pressure between two similar plates of finite
thickness kept at different temperatures in the case when the
dielectric permittivity of the plates depends on temperature. It is
suggested to consider the dielectric permittivity at two different
temperatures as the permittivities of two dissimilar bodies, thus
allowing to apply the theory of Casimir forces out of thermal
equilibrium developed earlier in the literature. Following this
approach, we show that, in addition to the equilibrium contribution
to the nonequilibrium Casimir pressure, a proper nonequilibrium
contribution arises for temperature-dependent dielectric
permittivities. Furthermore, the equilibrium contribution in this
case does not equal the mean of the equilibrium Casimir pressures
at the temperatures of the plates. As an application, the total
nonequilibrium Casimir pressure between two gold plates and between
two titanium plates is calculated as a function of the plate
thickness and their separation, using the Drude and the plasma
model. For plate separations ranging from 0.5 to $2~\mu$m, the
relative difference between the theoretical predictions for these
two models reaches 39\%. The proper nonequilibrium term may be as
large as 4\% of the magnitude of the total nonequilibrium pressure.
\end{abstract}

\maketitle

\section{Introduction}

Electromagnetic fluctuations and fluctuation-induced phenomena, such as the
Casimir and Casimir-Polder forces, are of widespread relevance and have gained
increasing interest due to advances in experimental techniques allowing for the
fabrication of various microstructures and their manipulation on the nanometer
scale \cite{1}. Different elements of a microstructure may be kept at different
temperatures, i.e., out of thermal equilibrium. This raises the question of
whether the standard Lifshitz theory describing the Casimir and Casimir-Polder
forces \cite{2,3} can be generalized to account for nonequilibrium conditions.

The extension of the Lifshitz formula to the case of two semispaces made of
absorbing materials at different temperatures was obtained in Ref.~\cite{4} in
the framework of fluctuational electrodynamics. For the special situation where
both semispaces are filled with the same material, it was shown that the
nonequilibrium Casimir pressure is given by the mean of the equilibrium Casimir
pressures at the temperatures of the two semispaces \cite{4}. The
Casimir-Polder force between an atom and a plate whose temperature is different
from that of the environment was considered in Refs.~\cite{5,6}. We note also
Ref.~\cite{7} where a theory of electromagnetic fluctuations has been used to
compute the radiative heat transfer between two parallel metallic surfaces kept
at different temperatures.

The general theory of Casimir and Casimir-Polder interactions in nonequilibrium
situations was developed in Refs.~\cite{8,9}. In the framework of this theory,
the Casimir pressure out of thermal equilibrium was presented as a sum of three
contributions: the mean of two equilibrium pressures at respective
temperatures, a proper nonequilibrium contribution and a term which is
independent of the separation between the plates. For two semispaces made of
identical materials it was confirmed that the Casimir pressure out of thermal
equilibrium reduces to the mean of two equilibrium pressures. This is also
true for two identical bodies of arbitrary shape placed symmetrically
with respect to a plane \cite{9}. Later, the theory of the Casimir force
out of equilibrium
was generalized to two or more arbitrarily shaped bodies consisting of
dissimilar materials  \cite{10,11,12,13,14,15,16,17}.
In Ref.~\cite{18} the possibility of a
nonequilibrium  repulsive force was proposed. Furthermore, the nonequilibrium
Casimir force was considered in connection with the effect of radiative heat
transfer, noncontact friction \cite{19,20,21,22}, and actuation of microdevices
\cite{22a}.

It is well known that the standard Lifshitz theory at thermal equilibrium faces
problems in describing the Casimir free energy and the pressure between two
metallic plates (see reviews \cite{23,24} and the monograph \cite{25}).
Specifically, it turns out that the results for the Casimir interaction
obtained by all high-precision experiments at separations below $1.1~\mu$m are
in conflict with theoretical predictions taking into account the relaxation
properties of conduction electrons \cite{26,27,28,29,30,31,32,33,34,35,36,37}.
If these relaxation properties are disregarded in computations, the Lifshitz
theory is brought to a very good agreement with the measurement data
\cite{26,27,28,29,30,31,32,33,34,35,36,37}. Even going beyond the usually
employed proximity force approximation by means of numerically exact
calculations does not help to resolve this discrepancy
\cite{Hartmann17,Hartmann18}. How the experimental observations can be brought
into agreement with the presence of relaxation processes for the conduction
electrons remains a puzzle \cite{38}.

An investigation of the Casimir force out of thermal equilibrium may shed more
light on the Casimir puzzle and point the way for its resolution.  This line of
attack was already used in Refs.~\cite{39,40} where the nonequilibrium Casimir
force between metallic test bodies was calculated with and without taking into
account relaxation properties of the conduction electrons and different
possibilities to measure it were proposed. Arguing that corrections due to the
temperature dependence of the dielectric permittivity would be small compared
to the effect of a temperature difference between the plates, the dielectric
permittivity was assumed to be temperature-independent.

In the present paper, we consider the Casimir pressure between two similar
metallic plates of finite thickness kept at different temperatures and taking into
account the temperature dependence of their dielectric permittivity. Although
the general theory of Refs.~\cite{8,9} does not explicitly account for such a
temperature dependence, it allows for different materials and therefore is
still applicable in our case. Then, one should consider the plate material at
two different temperatures as two dissimilar materials with different
dielectric permittivities depending only on frequency.

Following this approach, we demonstrate that a dependence of the dielectric
permittivity on temperature leads not only to quantitative, but also to
important qualitative effects. Specifically, we show that for two similar
metallic plates the nonequilibrium Casimir pressure does not reduce to the mean
of equilibrium pressures taken at the two different temperatures of the plates
as would be the case for bodies with temperature-independent dielectric
permittivity \cite{4,8,9}. Not only is this equilibrium contribution modified,
but also an additional proper nonequilibrium term contributes to the total
pressure.

As an application of the approach just outlined, we have performed numerical
computations of the Casimir pressure between two similar plates of finite
thickness made of either Au or Ti kept at different temperatures. The distance
between the plates varies from 0.5 to $2~\mu$m where we can restrict our
consideration to either the Drude or the plasma model \cite{25}. The plasma
frequency describing the high-frequency transparency of metals appears in both
models and can be taken as independent of temperature. In addition, the Drude
model includes relaxation properties of the conduction electrons where the
temperature dependence needs to be taken into account.

We have computed the magnitudes of the nonequilibrium Casimir pressure as a
function of separation for different plate thicknesses using either the Drude
or the plasma model. The corresponding results differ significantly and can be
discriminated experimentally. Furthermore, we determined the relative
difference between the standard and modified equilibrium contributions to the
nonequilibrium Casimir pressure. The magnitude of this quantity was found to be
less than 1\%. We have also computed the values of the proper nonequilibrium
contribution to the nonequilibrium Casimir pressure at different separations
and plate thicknesses. It is shown that this contribution can reach 4\% of the
magnitude of the total nonequilibrium pressure.

The paper is organized as follows. In Sec.~II the general formalism is presented
which allows to calculate the nonequilibrium Casimir pressure between two
similar plates with temperature-dependent dielectric permittivity. Section~III
contains our computational results for two similar plates made either of gold
or titanium. In Sec.~IV we will present our conclusions and discussion.

\section{General formalism for similar metallic plates out of thermal
equilibrium}

We consider the configuration of two parallel metallic plates of thickness $d$
in vacuum
consisting of the same material and separated by a distance $a$ as depicted in
Fig.~\ref{fg1}. The upper plate is kept at the temperature $T_1$ assumed to
equal the temperature of the environment while the lower plate has a different
temperature $T_2$. The plate material is characterized by a dielectric
permittivity $\ve(\omega, T)$ which in addition to its frequency dependence may
in general also vary with temperature. In this case, one can consider the two
plates as effectively consisting of different materials with dielectric
permittivities $\ve^{(1)}(\omega)=\ve(\omega, T_1)$ and
$\ve^{(2)}(\omega)=\ve(\omega, T_2)$, respectively.

Then, using the theory developed in Ref.~\cite{9}, the nonequilibrium Casimir
pressure on the lower plate can be expressed as the sum of two contributions
\begin{equation}
\pneq(a,T_1,T_2)={\tpeq}(a,T_1,T_2)+
\Delta\pneq(a,T_1,T_2).
\label{eq1}
\end{equation}
\noindent
Here, the quantity $\tpeq$ is a modification of the standard equilibrium
contribution of Refs.~\cite{4,9,18}
\begin{equation}
\bar{P}_{\rm eq}(a,T_1,T_2)=\frac{1}{2}\left[\peq(a,T_1)+\peq(a,T_2)\right]
\label{eq2}
\end{equation}
\noindent
equal to the mean of Casimir pressures calculated for the cases when both
plates are maintained either at the temperature $T_1$ or $T_2$.  The quantity
$\Delta\pneq$ is the proper nonequilibrium contribution to $\pneq$ which
vanishes for two similar plates possessing a temperature-independent dielectric
permittivity.

The explicit expression for  $\tpeq$ is given by
\begin{eqnarray}
&&
{\tpeq}(a,T_1,T_2)=-\frac{k_B}{2\pi}\left[
T_1\sum_{l=0}^{\infty}{\vphantom{\sum}}^{\prime}
\int_0^{\infty}q_l^{(1)}\kb d\kb \right.
\nonumber \\
&&
\times\sum_{\alpha}
\frac{R_{\alpha}(i\xi_l^{(1)}\!\!,\kb,T_1)
R_{\alpha}(i\xi_l^{(1)}\!\!,\kb,T_2)}{e^{2q_l^{(1)}a}
-R_{\alpha}(i\xi_l^{(1)}\!\!,\kb,T_1)R_{\alpha}(i\xi_l^{(1)}\!\!,\kb,T_2)}
\nonumber \\
&&~~
+T_2\sum_{l=0}^{\infty}{\vphantom{\sum}}^{\prime}
\int_0^{\infty}q_l^{(2)}\kb d\kb
\label{eq3} \\
&&\left.
\times\sum_{\alpha}
\frac{R_{\alpha}(i\xi_l^{(2)}\!\!,\kb,T_1)
R_{\alpha}(i\xi_l^{(2)}\!\!,\kb,T_2)}{e^{2q_l^{(2)}a}
-R_{\alpha}(i\xi_l^{(2)}\!\!,\kb,T_1)R_{\alpha}(i\xi_l^{(2)}\!\!,\kb,T_2)}
\right].
\nonumber
\end{eqnarray}
\noindent
Here, $k_B$ is the Boltzmann constant, $\kb$ is the magnitude of the wave vector projected
onto the plane defined by the plate surface, and the Matsubara frequencies $\xi_l^{(n)}$
with $n=1,\,2$ are defined as  $\xi_l^{(n)}=2\pi k_BT_nl/\hbar$,
$l=0,\,1,\,2,\,\ldots$.
The prime on the summation in $l$ divides the term with $l=0$ by 2 and $q_l^{(n)}$
is defined as
\begin{equation}
q_l^{(n)}=\sqrt{\kb^{\!\!\!2}+\frac{{\xi_l^{(n)}}^2}{c^2}}.
\label{eq4}
\end{equation}

The sums in  $\alpha$ are over the two polarizations  of electromagnetic
waves,  transverse magnetic
($\alpha={\rm TM}$) and transverse electric ($\alpha={\rm TE}$). The reflection coefficients
$R_{\alpha}$ on the plate of finite thickness $d$ are given by \cite{25}
\begin{equation}
R_{\alpha}(i\xi_l^{(n)}\!\!,\kb,T_m) =\frac{r_{\alpha}(i\xi_l^{(n)}\!\!,\kb,T_m)(1-
e^{-2dv_l^{(n,m)}})}{1-r_{\alpha}^2(i\xi_l^{(n)}\!\!,\kb,T_m)e^{-2dv_l^{(n,m)}}},
\label{eq5}
\end{equation}
\noindent
where $m=1,\,2$ denote the plates and the Fresnel coefficients for the reflection
at the surface of a semispace depend on the dielectric permittivity $\ve(\omega, T)$ as
\begin{equation}
\begin{aligned}
r_{\rm TM}(i\xi_l^{(n)}\!\!,\kb,T_m) &= \frac{\ve(i\xi_l^{(n)}\!\!,T_m)q_l^{(n)}-
v_l^{(n,m)}}{\ve(i\xi_l^{(n)}\!\!,T_m)q_l^{(n)}+v_l^{(n,m)}},
\\
r_{\rm TE}(i\xi_l^{(n)}\!\!,\kb,T_m) &=
\frac{q_l^{(n)}-v_l^{(n,m)}}{q_l^{(n)}+v_l^{(n,m)}}
\label{eq7}
\end{aligned}
\end{equation}
with
\begin{equation}
v_l^{(n,m)}=\sqrt{\kb^{\!\!\!2}+\ve(i\xi_l^{(n)}\!\!,T_m)\frac{{\xi_l^{(n)}}^2}{c^2}}.
\label{eq6}
\end{equation}

The expression (\ref{eq3}) is nothing else than the quantity $\peq$ of
Refs.~\cite{4,9,18} written for two plates made of materials with different
reflection coefficients
\begin{equation}
\begin{aligned}
R_{\alpha}^{(1)}(i\xi_l^{(n)}\!\!,\kb) &= R_{\alpha}(i\xi_l^{(n)}\!\!,\kb,T_1),
\\
R_{\alpha}^{(2)}(i\xi_l^{(n)}\!\!,\kb) &= R_{\alpha}(i\xi_l^{(n)}\!\!,\kb,T_2).
\label{eq8}
\end{aligned}
\end{equation}
\noindent
It should be pointed out, however, that (\ref{eq3}) reduces to the mean (\ref{eq2}) of
two equilibrium contributions at temperatures $T_1$ and $T_2$ only when the dielectric
permittivities are temperature-independent.

Following the same approach, the proper nonequilibrium contribution to the
nonequilibrium Casimir pressure (\ref{eq1}) can be written as follows \cite{9}
(see also Ref.~\cite{18})
\begin{widetext}
\begin{eqnarray}
&&
\Delta\pneq(a,T_1,T_2)=\frac{\hbar}{4\pi^2}\int_0^{\infty}\!\!\!\!d\omega
\left[\Theta(\omega,T_1)-\Theta(\omega,T_2)\right]
\nonumber \\
&&~~\times
\int_0^{\omega/c}\!\!\!d\kb \kb p\sum_{\alpha}
\frac{|R_{\alpha}(\omega,\kb,T_2)|^2-
|R_{\alpha}(\omega,\kb,T_1)|^2}{|D_{\alpha}(\omega,\kb,T_1,T_2)|^2}
\nonumber \\
&&~~
-\frac{\hbar}{2\pi^2}\int_0^{\infty}\!d\omega
\left[\Theta(\omega,T_1)-\Theta(\omega,T_2)\right]
\int_{\omega/c}^{\infty}\!d\kb \kb {\rm Im}p\,e^{-2a{\rm Im}p}
\label{eq9} \\
&&~~~~
\times\sum_{\alpha}\frac{{\rm Im}R_{\alpha}(\omega,\kb,T_1)
{\rm Re}R_{\alpha}(\omega,\kb,T_2)-{\rm Re}R_{\alpha}(\omega,\kb,T_1)
{\rm Im}R_{\alpha}(\omega,\kb,T_2)}{|D_{\alpha}(\omega,\kb,T_1,T_2)|^2},
\nonumber
\end{eqnarray}
\end{widetext}
where we have introduced the thermal photon population
\begin{equation}
\Theta(\omega,T_m)=\frac{1}{\exp\left(\frac{\hbar\omega}{k_BT_m}\right)-1},
\label{eq10}
\end{equation}
the component of the wave vector perpendicular to the plate surfaces
\begin{equation}
p=\sqrt{\frac{\omega^2}{c^2}-\kb^2},
\label{eq10a}
\end{equation}
and
\begin{equation}
D_{\alpha}(\omega,\kb,T_1,T_2)=1-R_{\alpha}(\omega,\kb,T_1)
R_{\alpha}(\omega,\kb,T_2)\,e^{2ipa}.
\label{eq11}
\end{equation}

The reflection coefficients for a plate of thickness $d$ appearing in (\ref{eq9})
as a function of real frequencies \cite{41}
\begin{equation}
R_{\alpha}(\omega,\kb,T_m) =\frac{r_{\alpha}(\omega,\kb,T_m)(1-
e^{2idu^{(m)}})}{1-r_{\alpha}^2(\omega,\kb,T_m)e^{2idu^{(m)}}},
\label{eq12}
\end{equation}
are expressed through the Fresnel coefficients
\begin{equation}
\begin{aligned}
r_{\rm TM}(\omega,\kb,T_m) &= \frac{\ve(\omega,T_m)p-
u^{(m)}}{\ve(\omega,T_m)p+u^{(m)}}, \\
r_{\rm TE}(\omega,\kb,T_m) &= \frac{p-u^{(m)}}{p+u^{(m)}}.
\label{eq14}
\end{aligned}
\end{equation}
where
\begin{equation}
u^{(m)}=\sqrt{\ve(\omega,T_m)\frac{\omega^2}{c^2}-\kb^{\!\!\!2}}.
\label{eq13}
\end{equation}
\noindent

Like in the case of the equilibrium contribution discussed above, Eq.~(\ref{eq9})
is obtained from the respective result of Ref.~\cite{9} for two dissimilar
plates by putting
\begin{equation}
\begin{aligned}
R_{\alpha}^{(1)}(\omega,\kb) &= R_{\alpha}(\omega,\kb,T_1), \\
R_{\alpha}^{(2)}(\omega,\kb) &= R_{\alpha}(\omega,\kb,T_2).
\label{eq15}
\end{aligned}
\end{equation}
If the dielectric permittivity is temperature-independent, one obtains
$R_{\alpha}^{(1)}=R_{\alpha}^{(2)}$ for two similar plates and  Eq.~(\ref{eq9})
leads to $\Delta\pneq=0$ \cite{4,9,18}. This is not the case, however, for two
similar plates described by a temperature-dependent dielectric permittivity.

As was mentioned above, the nonequilibrium Casimir pressure (\ref{eq1}) acts on the
lower plate which is kept at the temperature $T_2$ different from the temperature
of the environment. Then, the pressure on the upper plate kept at the
temperature of the environment $T_1$ contains an additional contribution from
the environmental black-body radiation pressure\cite{18,40}
\begin{equation}
\widehat{P}_{\rm neq}(a,T_1,T_2)=\pneq(a,T_1,T_2)+\frac{2\sigma}{3c}(T_2^4-T_1^4),
\label{eq16}
\end{equation}
\noindent
where $\sigma$ is the Stefan-Boltzmann constant and $\pneq$ is defined through
Eqs.~(\ref{eq1}), (\ref{eq3}) and (\ref{eq9}).

In the next section we will make use of the formalism just presented in order
to numerically determine the nonequilibrium Casimir pressure and its different
contributions for two parallel metallic plates of finite thickness.

\section{Computational results for gold and titanium plates}

To demonstrate physical consequences of the formalism presented in Sec.~II, we
consider two similar parallel metallic plates kept at temperatures $T_1=300\,$K
and $T_2=500\,$K and separated by distances $a\geq 0.5\,\mu$m. At such
separations, the Casimir pressure is largely determined by the contribution of
conduction electrons to the dielectric permittivity of a metal whereas the
contribution of core electrons results only in minor corrections \cite{25}.
The chosen temperatures do not appear as unrealistic in view of existing
experiments \cite{Obrecht07}.

The most frequently used and well confirmed dielectric permittivity describing
the conduction electrons in metals including their relaxation properties is
given by the Drude model
\begin{equation}
\ve_D(\omega,T_m)=1-\frac{\omega_p^2}{\omega[\omega+i\gamma(T_m)]},
\label{eq17}
\end{equation}
\noindent
where $\omega_p$ is the plasma frequency. This dielectric permittivity is
temperature-dependent through the temperature dependence of the
relaxation parameter $ \gamma(T_m)$.
Here, $T_m$ is the temperature of a metallic plate in local thermal equilibrium
\cite{9} ($T_1$ and $T_2$ for upper and lower plates, respectively).
At the purely imaginary
Matsubara frequencies $i\xi_l^{(n)}$ the permittivity (\ref{eq17}) takes the form
\begin{equation}
\ve_D(i\xi_l^{(n)}, T_m)=1+\frac{\omega_p^2}{\xi_l^{(n)}[\xi_l^{(n)}+\gamma(T_m)]}.
\label{eq18}
\end{equation}

As discussed in Sec.~I, theoretical predictions of the Lifshitz theory using the
Drude model (or the optical data of metals extrapolated by means of the Drude
model down to zero frequency) are excluded by high-precision measurements
of the Casimir force. For reasons unknown to us the measurement data
are in agreement with the theoretical predictions disregarding the relaxation
properties of conduction electrons.

Setting the relaxation parameter $\gamma$ in Eqs.~(\ref{eq17}) and (\ref{eq18})
strictly to zero \cite{23,Ingold09}, the dissipative Drude model is replaced by
the dissipationless plasma model
\begin{equation}
\ve_p(\omega)=1-\frac{\omega_p^2}{\omega^2},\quad
\ve_p(i\xi_l^{(n)})=1+\frac{\omega_p^2}{{\xi_l^{(n)}}^2}.
\label{eq19}
\end{equation}

An important difference between the plasma model and the Drude model consists
in the fact that the Drude model is temperature-dependent while the plasma model
is not. Below, we perform numerical computations of the nonequilibrium Casimir
pressure using both models.
For the plate material, we choose either gold or titanium which possess rather
different Drude parameters as listed in Table~\ref{tableI}.

Computations of the magnitude of the nonequilibrium Casimir pressure
(\ref{eq1}) are performed using Eqs.~(\ref{eq3})--(\ref{eq6}) and
(\ref{eq9})--(\ref{eq13}). While Eq.~(\ref{eq3}) is expressed in
terms of a summation over the Matsubara frequencies,  Eq.~(\ref{eq9})
is presented in terms of integrals of oscillating integrands over real
frequencies. It is common knowledge that a problem of numerical
convergence in the Lifshitz formula at real
frequencies refers to only the total
Casimir pressure and is unrelated to the thermal correction \cite{45a,45b}
due to the presence of the factor $\Theta(\omega,T)$.
In our case Eq.~(\ref{eq9}) is very much like to the thermal correction,
and the convergence is ensured by a presence of the factors
$\Theta(\omega,T_m)$.

The computational results as function of the
separation between the plates are shown in Fig.~\ref{fg2}(a) for Au and in
Fig.~\ref{fg2}(b) for Ti. The top line in Fig.~\ref{fg2}(a) is computed
using the plasma model (\ref{eq19}) and the bottom line  using the Drude
model (\ref{eq17}) and (\ref{eq18}). Each of these lines is actually a
superposition of two lines computed for plates of $d=20~$nm and $1~\mu$m
thickness. Note that even the plates of minimum thickness considered
($d=20~$nm)  allow theoretical description in terms of the bulk
dielectric permittivity. For instance, according to results of
Refs.~\cite{45c,45d}, the Casimir pressure can be calculated using the
bulk dielectric permittivity if the plates made of Au consist of more than
30 atomic layers, i.e., are of more than 7~nm thickness.
Approximately the same results are expected for Ti plates.

The pairs of lines labeled 1 and 2 in Fig.~\ref{fg2}(b) are
computed for plates of $d=20\,$nm and $1\,\mu$m thickness, respectively. The
upper line  in each pair was obtained using the plasma model while the
lower line  corresponds to the Drude model. As can be seen in
Fig.~\ref{fg2}, the value of $\pneq$ essentially depends on the model employed
for the dielectric permittivity. While for Au plates there is basically no
dependence on the plate thickness $d$ for the values chosen, the dependence is
more pronounced for Ti plates.

To clearly demonstrate the difference in the values of $\pneq$ caused by the use
of different permittivity models for plates of different thickness, we have also
computed the quantity $\pneq/P_0$ where
\begin{equation}
P_0(a)=-\frac{\pi^2}{240}\,\frac{\hbar c}{a^4}
\label{eq21}
\end{equation}
is the Casimir pressure between two ideal metal plates at zero temperature.
The computational results are shown on a linear scale as function of the
separation $a$ for Au in Fig.~\ref{fg3}(a) and Ti in Fig.~\ref{fg3}(b).  In
Fig.~\ref{fg3}(a), the top and bottom  pairs of solid lines are
computed using the plasma and Drude models, respectively. In each of these
pairs, the lower solid line corresponds to $d=20~$nm and the upper solid line to
$d=1~\mu$m.  In Fig.~\ref{fg3}(b), the pairs of solid lines labeled 1 and 2 are
computed for the plates with $d=20~$nm and $1~\mu$m thickness, respectively.
In each pair, the upper solid line  is obtained using the plasma model and the
lower solid line  using the Drude model. In both, Figs.~\ref{fg3}(a) and
\ref{fg3}(b), the dashed lines are computed using the Drude model with the
temperature-independent relaxation parameter fixed at its value at 300\,K.

As Fig.~\ref{fg3} shows, the theoretical predictions based on the Drude and
plasma models differ more than sufficiently in order to discriminate between
them experimentally. For instance, for Au plates separated by
$a=0.5\,\mu\text{m}, 1\,\mu\text{m}$ and $2\,\mu\text{m}$ the relative
differences between the nonequilibrium Casimir pressures computed using the
Drude and plasma models are equal to 12\%, 22\%, and 39\% for plates of
20~nm thickness, respectively. For plates of $1~\mu$m thickness, the
respective relative differences are 11\%, 21\%, and 38\%, i.e., the dependence
on the thickness is very minor. For Ti plates approximately the same results
are obtained.  Furthermore, from Fig.~\ref{fg3} one can again see that the
thickness makes a greater impact on $\pneq$ for Ti plates than for Au plates.
The differences between the dashed lines and the neighboring solid lines
illustrate the effect originating from the temperature dependence of the
dielectric permittivity. Below we consider them in more detail.

According to Sec.~II, a dependence of the dielectric permittivity on
temperature makes an impact on both the equilibrium and nonequilibrium
contributions to the total nonequilibrium Casimir pressure (\ref{eq1}). We
begin with the first of them and consider the relative difference
\begin{equation}
\delta\peq(a,T_1,T_2)=\frac{\tpeq(a,T_1,T_2)-
\bar{P}_{\rm eq}(a,T_1,T_2)}{\bar{P}_{\rm eq}(a,T_1,T_2)}.
\label{eq22}
\end{equation}
between $\tpeq$ and $\bar{P}_{\rm eq}$ defined in Eqs.~(\ref{eq3}) and
(\ref{eq2}), respectively, The computations are made for the dielectric
permittivity of the Drude model. For the plasma model, it is evident that
$\delta\peq\equiv 0$.

In Fig.~\ref{fg4} the computational results for $\delta\peq$ in percent are
shown as function of the plate separation by the four lines from top to
bottom for the cases of Au plates with $d=1~\mu$m and 20~nm and Ti plates
with $d=1~\mu$m and 20~nm, respectively.
As can be seen in Fig.~\ref{fg4}, for Au plates the magnitudes of the relative
difference between $\tpeq$ and $\bar{P}_{\rm eq}$ do not exceed 0.1\%
independently of the values of plate thickness and separation distance.
For Ti plates the impact of the plate thickness on $\delta\peq$ is more
pronounced, but even in this case the magnitudes of $\delta\peq$ at different
separations do not reach 0.7\% for plates of 20~nm thickness.
One can conclude that an impact of the temperature dependence of the dielectric
permittivity of a metal on the equilibrium contribution to the nonequilibrium
Casimir pressure, although of interest theoretically, is negligibly small from
the experimental point of view.

Now we consider the role of the second contribution to the nonequilibrium
Casimir pressure (\ref{eq1}). For similar plates it arises only if the
dielectric permittivity of the plate material is temperature-dependent.  In
Fig.~\ref{fg5} the ratios $\Delta\pneq/\pneq$ are shown as functions of the
plate separation by the two lines from top to bottom for (a) Au plates of
$d=1~\mu$m and 20~nm thickness and (b) Ti plates of $d=20~$nm and $1~\mu$m
thickness. Computations are performed using the dielectric permittivity of the
Drude model (\ref{eq17}) and (\ref{eq18}).  For the plasma model one has
$\Delta\pneq\equiv 0$.

As can be seen from Fig.~\ref{fg5}(a), for Au plates of 20~nm thickness the proper
nonequilibrium contribution $\Delta\pneq$ reaches 4\% of the magnitude of the
total nonequilibrium Casimir pressure at a separation of $a=2~\mu$m between the
plates. While $\pneq$ is negative, $\Delta\pneq$ is positive, i.e., it makes a
repulsive contribution over the range of separations considered.  For Ti plates
of $1~\mu$m thickness, $\Delta\pneq$ reaches 3\% of $|\pneq|$ at $a=2~\mu$m. It
should be noted that for Ti plates of 20~nm thickness at less than $1.19~\mu$m
separation, $\Delta\pneq$ changes its sign and becomes negative, i.e. it has the
same sign as $\pneq$. Thus, at sufficiently small separations between the
plates it contributes to the standard Casimir attraction.

Physically the results are explained by the difference in penetration depth
$c/\omega_p$ for electromagnetic fluctuations in Au and Ti of 22\,nm and
78.9\,nm, respectively. While for Au the thinnest plate considered is almost of
the same thickness as the penetration depth, for Ti the penetration depth
exceeds the thickness of the thinnest plate considered by almost a factor of
four.

We can conclude that for similar plates of sufficient thickness and with
temperature-dependent dielectric permittivity the proper nonequilibrium
contribution $\Delta\pneq$ can amount to several percent of the total pressure.
It should thus be taken into account when comparing experiment and theory.

For completeness we have also calculated the proper nonequilibrium
contribution to the Casimir pressure $\Delta\pneq$ as a function of plate
thickness $d$.  The computational results are presented in Fig.~\ref{fg6}
by the top and bottom pairs of lines obtained at separations between the
plates $a=0.5$ and $1~\mu$m, respectively. The upper line in each pair is
plotted for Au plates and the lower line is for Ti plates. As can be seen in
Fig.~\ref{fg6}, the quantity $\Delta\pneq$ reaches its maximum value for
plate thicknesses $d$ approximately equal to twice the penetration depth
of electromagnetic fluctuations in Au and Ti. For Au, the quantity
$\Delta\pneq$ remains positive for any plate thickness in the range from
20~nm to $1~\mu$m, i.e., corresponds to a repulsive contribution to the
Casimir force.

From Fig.~\ref{fg6} it can be seen that for Ti plates of thickness $d\geq
22.1\,$nm at a distance $a=0.5\,\mu$m, $\Delta\pneq$ is also positive. For
thinner Ti plates, however, $\Delta\pneq$ changes sign and becomes negative.
For instance, for Ti plates of thickness $d=20\,$nm we have
$\Delta\pneq=-3.04\,\mu$Pa. For Ti plates spaced at $a=1\,\mu$m we have
$\Delta\pneq>0$ for plates of thickness $d\geq 21.1\,$nm. For thinner plates
$\Delta\pneq<0$. Specifically, for plates with $d=20\,$nm one finds
$\Delta\pneq=-0.17\,\mu$Pa. We underline that for similar Ti plates
of some specific value of the plate thickness depending on the plate distance
the proper nonequilibrium contribution to the Casimir pressure vanishes. In
this case the total nonequilibrium pressure (\ref{eq1}) is reduced to the
equilibrium contribution $\tpeq$ alone.

At the end of this section, we recall that all the above results are formulated
for the total pressure $\pneq$ on the lower plate. According to
Eq.~(\ref{eq16}), the total pressure on the upper plate
$\widehat{P}_{\rm neq}$ is obtained by an additional constant term.
For the temperatures used in our computations ($T_1=300\,$K and $T_2=500\,$K) this
contribution amounts to
\begin{equation}
\frac{2\sigma}{3c}(T_2^4-T_1^4)=6.86~\mu\mbox{Pa}.
\label{eq23}
\end{equation}
From a comparison with Fig.~\ref{fg6}, it is seen that the constant term is
of approximately the same size as $\Delta\pneq$ for plates at a separation of
$a=1~\mu$m.

\section{Conclusions and discussion}

In this paper, we have applied the theory of Casimir forces
out of thermal equilibrium to the case of two similar plates
possessing a temperature-dependent dielectric permittivity.
This problem, not studied so far, is of importance
in connection with the so-called Casimir puzzle awaiting its
resolution for already twenty years. It is relevant also for
developing applications of the Casimir force in various
microdevices.

We suggested here that if two similar plates possessing a
temperature-dependent dielectric permittivity are kept at different
temperatures, they can be considered as dissimilar bodies with
different dielectric permittivities. This approach allows the
application of a theory developed earlier in the literature
\cite{8,9}. According to our results and in contrast to what was
believed so far for two identical bodies, the nonequilibrium Casimir
pressure for two similar metallic plates possessing a
temperature-dependent dielectric permittivity does not reduce to an
equilibrium contribution alone but contains a nonzero proper
nonequilibrium contribution as well. Furthermore, in this case the
equilibrium contribution undergoes a modification and is not equal to
the mean of two standard equilibrium pressures calculated for
pairs of plates kept at two different temperatures.  The configuration
of two dissimilar plates possessing a temperature-dependent
dielectric permittivity can be treated in an analogous way.

The developed formalism was applied to metallic plates of equal finite
thickness made of either Au or Ti and spaced at sufficiently large
separations so that an influence of core electrons on the Casimir
force is negligibly small. We have computed the total nonequilibrium
Casimir pressure and both the equilibrium and proper nonequilibrium
contributions to it at different separations and plate thicknesses
using the dielectric permittivities of the temperature-dependent Drude
model and the temperature-independent plasma model. It is shown that
the relative difference between the total nonequilibrium pressures
computed using the Drude and plasma models reaches 39\% when the
separation distance between the plates varies from 0.5 to $2~\mu$m.
The magnitude of the relative difference between the modified and
standard equilibrium contributions to the total pressure is shown to
be less than 1\%. However, according to our results,  the proper
nonequilibrium contribution to the total pressure between two similar
plates computed using the Drude model can reach 4\% of the magnitude
of the total pressure for Au plates, i.e., it should be taken into
account in the comparison of experiment and theory. It is interesting
that for Ti plates of a certain thickness depending on separation (as
an example, for plates separated by $0.5~\mu$m, this
thickness equals 22.1\,nm) the proper nonequilibrium contribution
to the total pressure vanishes and changes sign to yield an attractive
contribution for thinner plates.

To conclude, according to the above results, the nonequilibrium
Casimir pressure between similar plates made of a material
possessing a temperature-dependent dielectric permittivity
demonstrates not only quantitative, but also qualitative
effects, which are lacking for bodies described by a
temperature-independent permittivity and may be of interest for
future experiments and technological applications.
These effects could be observed using the experimental setup described
in Refs.~\cite{40,48}, where the upper plate plays the role of a sensor
suspended by the springs and the lower plate is mounted on a
piezo-electric stage. Taking into account that an immediate application of
the above computational results requires the vacuum conditions outside
of both plates, their thickness should be chosen sufficiently large
(typically more than $0.5~\mu$m) in order that a material underlying the
lower plate will not impact on the nonequilibrium pressure. One could
also suggest other types of experimental setups suitable for observation
of the proposed effects.

For the future it is planned to explore situations where the effects described in
this paper are more pronounced, e.g. for materials for which the dielectric
permittivity is more sensitive to changes in temperature.
The most interesting candidate for such a study is graphene, leading to
a large thermal effect in the Casimir interaction at short separations
\cite{49,50}.

\section*{Acknowledgments}
The work of G.L.K.~and V.M.M.~was partially supported by the
Peter the Great Saint Petersburg Polytechnic University in the framework of the
Program ``5--100--2020".
V.M.M.~was partially funded by the Russian Foundation for Basic
Research, Grant No. 19-02-00453 A. His work was also partially
supported by the Russian Government Program of Competitive Growth
of Kazan Federal University.
G.L.K.~and V.M.M.~are grateful to the Department
of Physics of the University of Augsburg, where part of this
work was performed, for kind hospitality. They acknowledge
helpful discussions with M.~Antezza and V.~B.~Svetovoy.

\begin{table}
\caption{\label{tableI} Plasma frequency $\omega_p$ and relaxation parameter $\gamma$
	at two different temperatures for gold and titanium \cite{42}.}
\begin{ruledtabular}
\begin{tabular}{lrrr}
metal & $\hbar\omega_p$ & $\hbar\gamma(300\,\text K)$ & $\hbar\gamma(500\,\text K)$ \\ \hline
Au    & 9.0\,\text{eV}  &  35\,\text{meV}             &  58\,\text{meV} \\
Ti    & 2.51\,\text{eV} &  47\,\text{meV}             &  78\,\text{meV} \\
\end{tabular}
\end{ruledtabular}
\end{table}
\begin{widetext}
\begin{center}
\begin{figure}[b]
\vspace*{-0cm}
\centerline{\hspace*{0.3cm}
\includegraphics{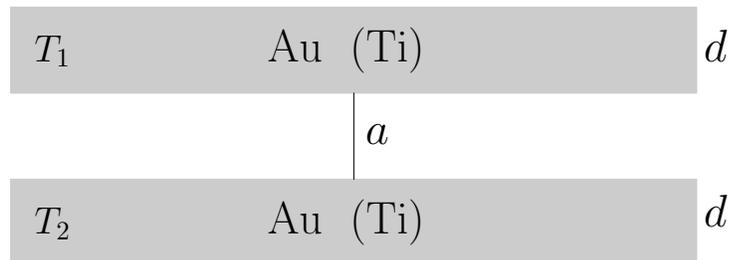}
}
\vspace*{-17.cm}
\caption{\label{fg1}
The configuration of two parallel plates of thickness $d$ which
are kept at different temperatures $T_1$ (as in the environment)
and $T_2$.
}
\end{figure}
\begin{figure}[b]
\vspace*{-3cm}
\centerline{\hspace*{0.3cm}
\includegraphics{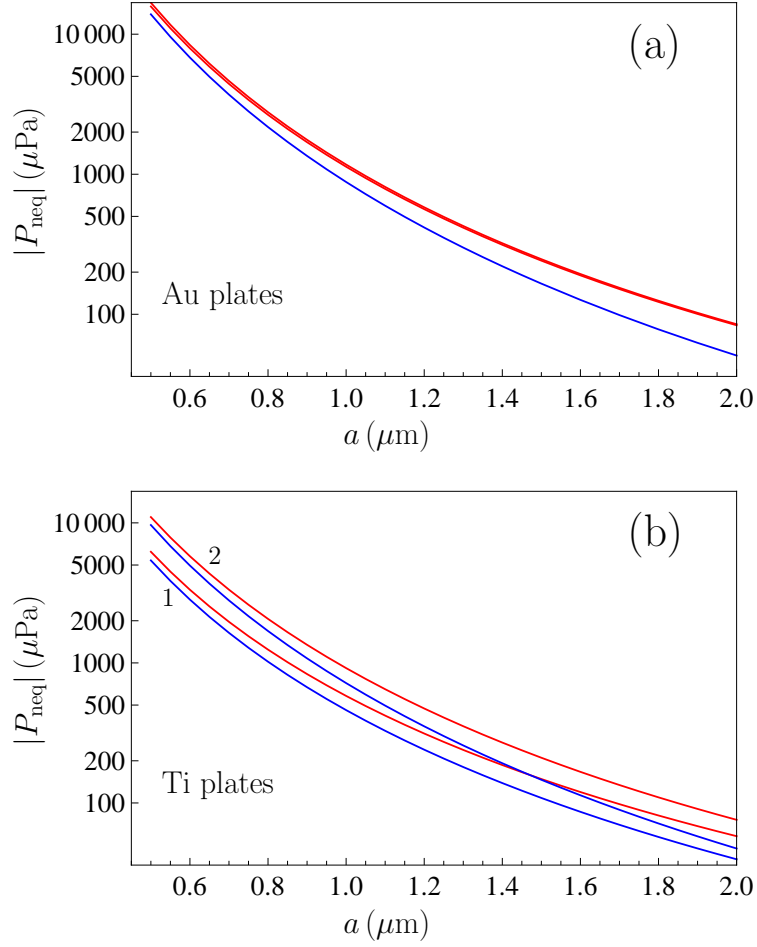}
}
\vspace*{-13.cm}
\caption{\label{fg2}
The magnitudes of the nonequilibrium Casimir pressure are shown
as functions of separation (a) for Au plates by the top
and bottom  lines computed using the plasma and Drude
models, respectively, and (b) for Ti plates by the pairs of lines labeled
1 and 2 computed for the plates of $d=20$~nm and $1~\mu$m thickness,
respectively, where in each pair the upper and lower lines are
obtained using the plasma and Drude models, respectively.
}
\end{figure}
\begin{figure}[b]
\vspace*{-3cm}
\centerline{\hspace*{0.3cm}
\includegraphics{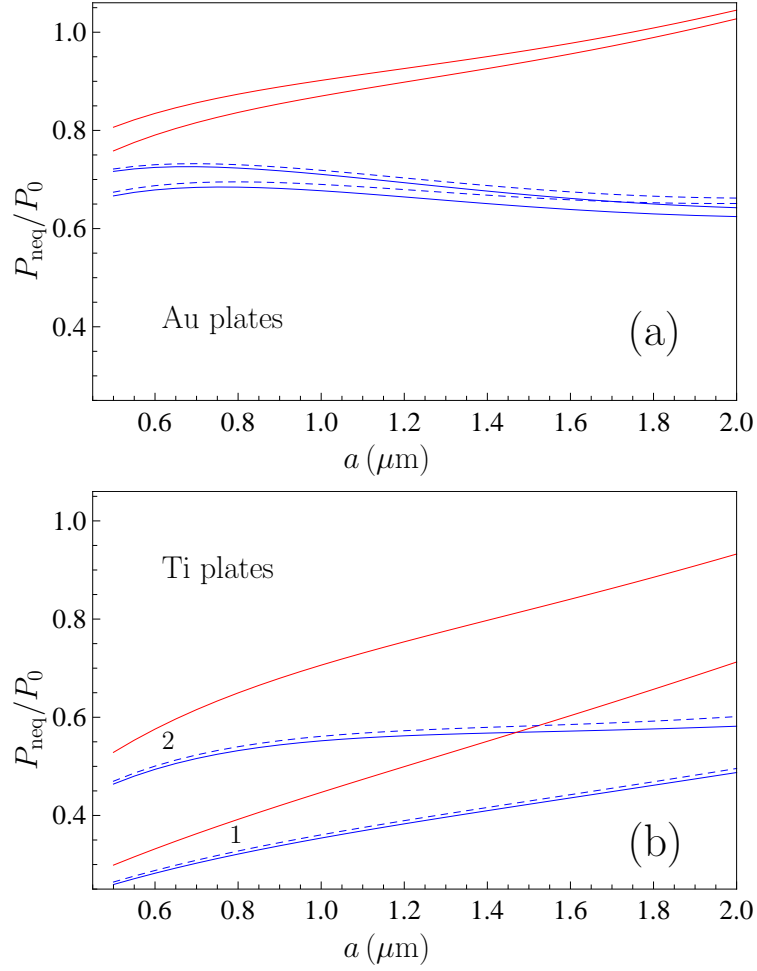}
}
\vspace*{-13.cm}
\caption{\label{fg3}
The nonequilibrium Casimir pressures normalized to the Casimir
pressure between two ideal metal plates at zero temperature are
shown as functions of separation (a) for Au plates by the top and
bottom pairs of solid lines computed using the plasma and Drude
models, respectively, where in each pair the lower and upper solid
lines are for a plate thickness of $d=20$~nm and $1~\mu$m,
respectively, and (b) for Ti plates by the pairs of solid lines
labeled 1 and 2 computed for the plate thickness of $d=20$~nm and
$1~\mu$m, respectively, where in each pair the upper and lower
solid lines are obtained using the plasma and Drude models,
respectively. In all cases the dashed lines are computed by means
of the Drude model with the temperature-independent relaxation
parameter taken at 300\,K.
}
\end{figure}
\begin{figure}[b]
\vspace*{-0cm}
\centerline{\hspace*{0.3cm}
\includegraphics{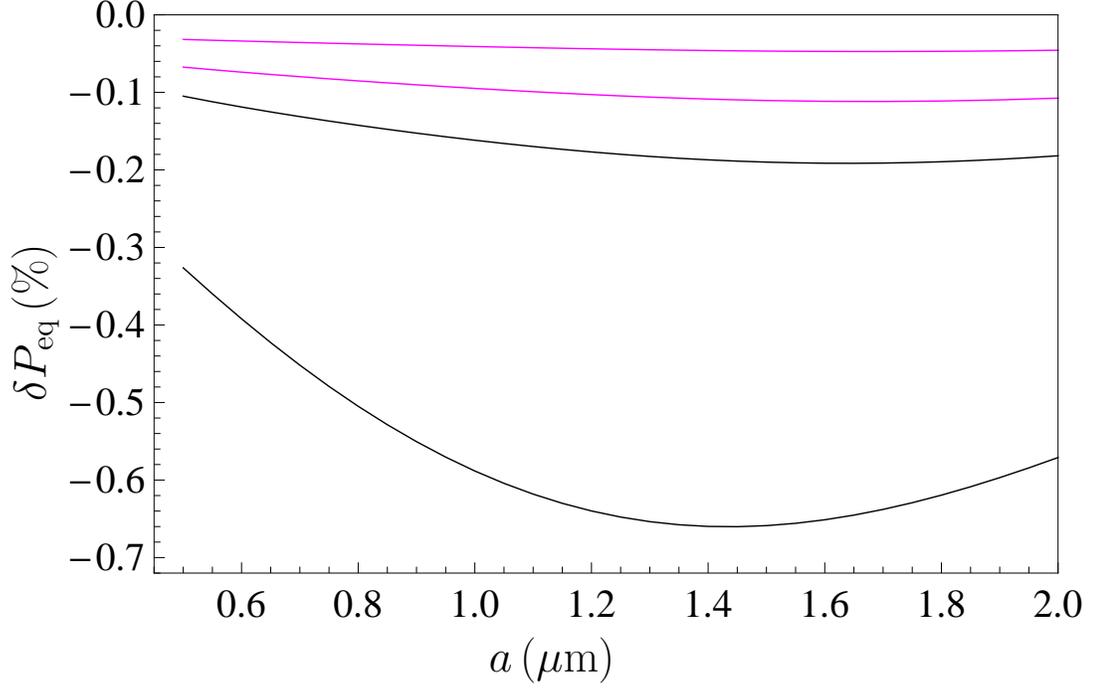}
}
\vspace*{-17.cm}
\caption{\label{fg4}
The relative difference between the modified and standard
equilibrium contributions to the nonequilibrium Casimir pressure
between two similar metallic plates described by the Drude model
with temperature-dependent relaxation parameter as a function of
separation is shown in percent by the four lines from
top to bottom for Au plates with $d=1~\mu$m and 20~nm thickness
and for Ti plates with $d=1~\mu$m and 20~nm, respectively.
}
\end{figure}
\begin{figure}[b]
\vspace*{-3cm}
\centerline{\hspace*{0.3cm}
\includegraphics{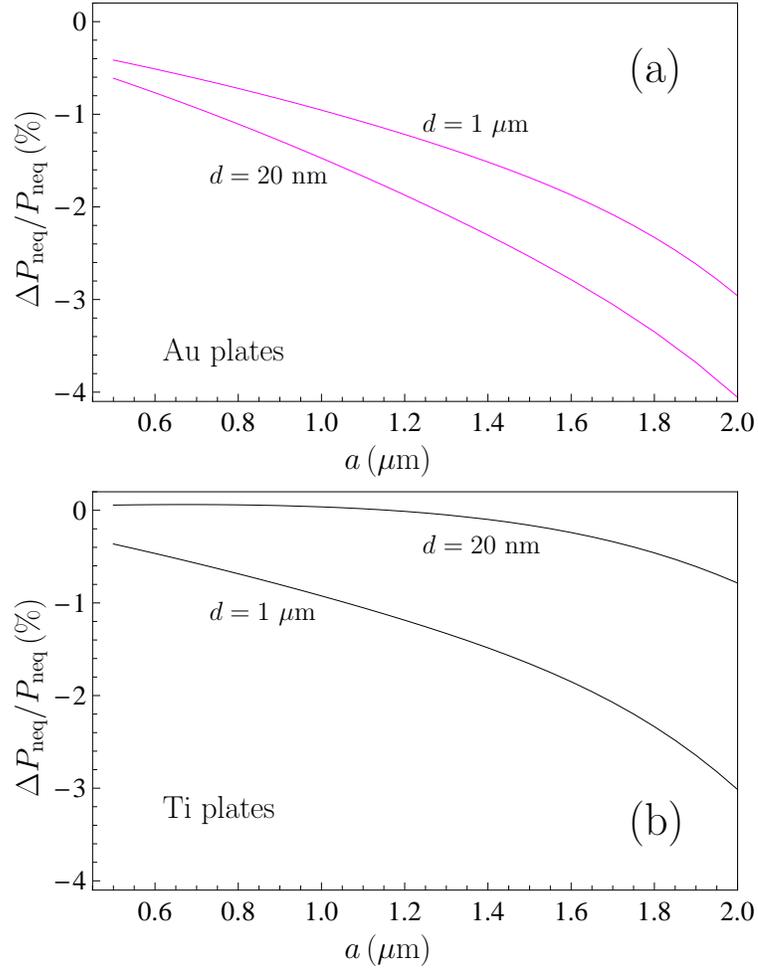}
}
\vspace*{-13.cm}
\caption{\label{fg5}
The ratios of the proper nonequilibrium contribution and the total
Casimir nonequilibrium pressure are shown in percent as function
of the separation by the two solid lines from top to bottom for
(a) Au plates with $d=1~\mu$m and 20~nm thickness, respectively, and
(b) Ti plates with $d=20$~nm and $1~\mu$m, respectively.
}
\end{figure}
\begin{figure}[b]
\vspace*{-0cm}
\centerline{\hspace*{0.3cm}
\includegraphics{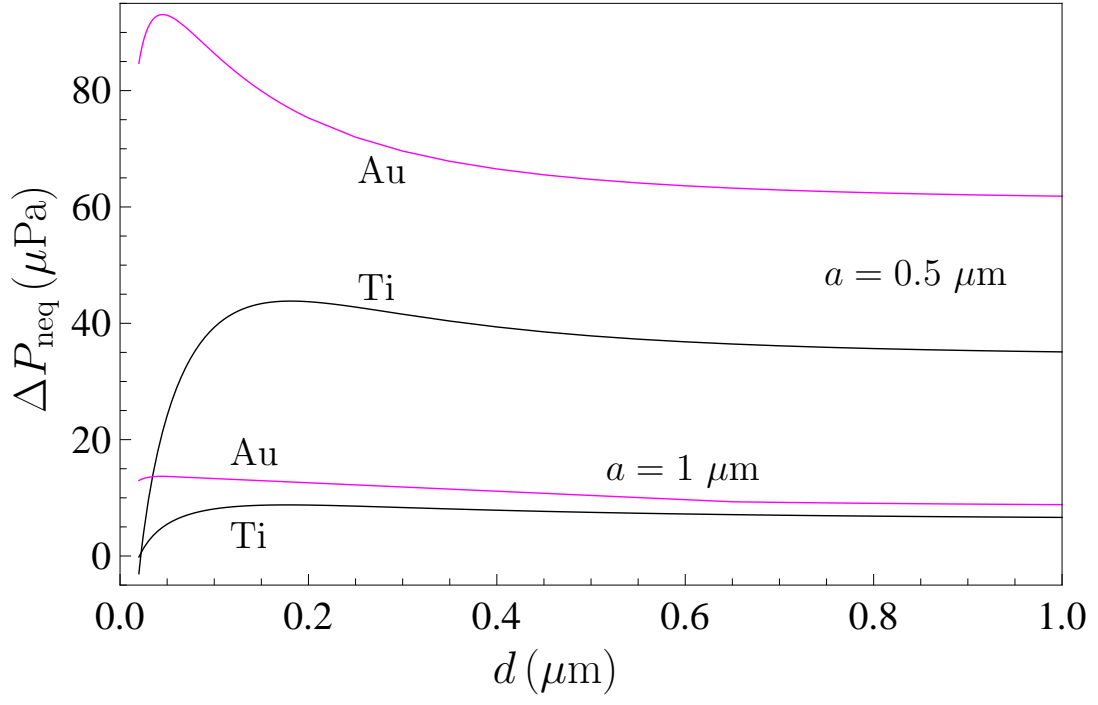}
}
\vspace*{-17.cm}
\caption{\label{fg6}
The proper nonequilibrium contribution to the Casimir pressure is
shown as a function of plate thickness by the top and bottom pairs
of lines computed at separations $a=0.5$ and $1~\mu$m between the
plates, respectively. In each pair, the upper line corresponds to
Au plates while the lower line is for Ti plates.
}
\end{figure}
\end{center}
\end{widetext}
\end{document}